\documentclass[pra,preprint,showpacs]{revtex4}
\usepackage[centertags]{amsmath}
\usepackage{amsfonts}
\usepackage{amssymb}
\usepackage{amsthm} 
\usepackage{newlfont}
\usepackage{epsfig}
\usepackage{amscd}
\usepackage{graphicx}
\usepackage{epsfig}
\usepackage{amscd}

\newcommand{\beq}{\begin{equation}}
\newcommand{\eeq}{\end{equation}}
\newcommand{\ba}{\begin{array}}
\newcommand{\ea}{\end{array}}
\newcommand{\bea}{\begin{eqnarray}}
\newcommand{\eea}{\end{eqnarray}}
\begin{document}

\begin{center}
{\large \sc \bf {Quantum correlations in different density matrix representations of spin-1/2 open chain}
}

\vskip 15pt

{\large 
 E.B.Fel'dman and A.I.~Zenchuk 
}

\vskip 8pt

{\it Institute of Problems of Chemical Physics, RAS,
Chernogolovka, Moscow reg., 142432, Russia, e-mail:   zenchuk@itp.ac.ru
 } 
\end{center}


\begin{abstract}
We consider quantum correlations in a spin-1/2 open chain of $N$ nodes with the XY Hamiltonian using different bases for the density matrix representation and the initial state with a single  polarized node.
These bases of our choice are following: (i)  the basis of eigenvectors of the fermion operators; this basis   appears naturally through the Jordan-Wigner transformation (this representation of the  density matrix   is  referred to  as the 
$\beta$-representation), (ii) its  Fourier representation ($c$-representation of the density matrix) and (iii) the basis of eigenvectors of the operators $I_{jz}$ (the $z$-projection of the $j$th spin, $j=1,\dots,N$). Although for the short chains (a few nodes) the qualitative behavior of the entanglement and the discord are very similar (the difference is quantitative), this is not valid for longer chains ($N\gtrsim 10$). In this case, there are qualitative and quantitative  distinctions between  the entanglement and the discord  in all three cases. We underline three most important features: (i) the  quantum discord is static in the $\beta$-representation, where the entanglement is identical to zero; (ii) in the $c$-representation, the concurrence may be non-zero only between the nearest neighbors (with a single exception), while the discord is nonzero  between any two nodes; (iii) there is so-called ''echo'' in the evolution of the discord, which is not observed in the evolution of the concurrence.
Using different bases, we may choose the preferable behavior of quantum correlations  which allows a given quantum system to be more flexible in applications.
\end{abstract}

\maketitle

\section{Introduction}
\label{Sec:Introduction}

The  advantage of quantum correlations in comparison with classical ones rises a fast development of quantum information and communication devices.  However, the problem of identifying of  quantum  correlations in a given system is not resolved yet. 
First, the so-called  quantum entanglement \cite{Werner,HWootters,P,AFOV,DPF} was considered as a proper measure of quantum correlations. Later on, the quantum discord  was accepted  as a more adequate  measure of quantum correlations
\cite{Z0,HV,OZ,Z}. In particular, the states with zero entanglement can reveal non-zero discord. Therefore it was noted \cite{DSC} that namely discord might be responsible for advantages of quantum information devices, in particular, for  quantum speed-up. In general, the calculation of discord is very cumbersome optimization problem. In spite of  intensive study of discord 
only very special cases have been treated analytically  \cite{L,ARA,Xu}. 
Nevertheless, namely these cases  correspond to the reduced binary density matrix in spin-1/2 chains governed by  different Hamiltonians with either the thermal equilibrium initial state \cite{G,KZ} or the  initial state with a single polarized node.

A problem of a proper initial state is one of the fundamental problems in study of  quntum correlations in different physical systems because of the technical difficulties of realization of  a particular state.
Most popular is  so-called thermal equilibrium initial state \cite{G}, which is most simple for realization. However, the initial state with a single exited node is more relative, for instance, in quantum communication lines \cite{Bose,CDEL,ACDE,KS,GKMT,FKZ}. The  state with a single polarized node has been produced experimentally \cite{ZME}. The evolution of quantum systems  with this initial state  at high temperatures has been studied, for instance, in \cite{FBE}, where the quantum echo has been found.

Nevertheless, the problem of identification of quantum correlations is not resolved yet. In particular, the measure of quantum correlations depends on the basis which is taken for the density matrix representation. The reason is that, considering different bases, we involve  different types of  ''virtual particles''. A possible way to avoid this ambiguity is suggested in ref.\cite{Z_a}, where the unitary invariant discord is introduced. This measure  takes into account correlations among all possible ''virtual particles''. 

In this paper we consider the  problem of preferable ''virtual particles'' (or the preferable basis of the matrix representation) from a different viewpoint. Instead of collecting the correlations among all ''virtual particles'' (like in the unitary invariant discord \cite{Z_a}) we consider separately the quantum correlations among three types of particles, namely, among the fermions, which appear in a spin-1/2 system with the nearest neighbor interaction under the Jourdan-Wigner transformation \cite{JW}  (we call them as the $\beta$-fermions), among the fermions which are the  Fourier representations of the $\beta$-fermions (the $c$-fermions),  and among the spin-1/2 particles in the basis of eigenvectors of the operators $I_{jz}$ ($z$-projection of the $j$th spin, $j=1,\dots, N$). Each type of 
particles corresponds to the proper basis of the density matrix representation. We show, that the choice of the basis is a significant (may be even dominant) factor in identification of quantum correlations.

This paper is organized as follows.
The evolution of the spin-1/2 open chain of $N$ nodes under the XY Hamiltonian with the nearest neighbor interactions is derived in Sec.\ref{Section:chain} using the Jourdan-Wigner transformation. The general formulas for the discord and the concurrence
are discussed in Sec.\ref{Section:discord}. The comparison of the discord and the concurrence in different bases of the density matrix representation is given in Sec.\ref{Section:comparison}. The summary of our results is represented  in Sec.\ref{Section:conclusions}.


\section{Spin-1/2 chains with a single initially  polarized node}
\label{Section:chain}

In this paper we study the quantum correlations in the one-dimensional open spin-1/2 chain of $N$ nodes governed by the 
XY Hamiltonian in  the approximation of  nearest neighbor interactions,
\begin{eqnarray}\label{XY}
H=\omega_0 \sum_{i=1}^N I_{iz} + D \sum_{i=1}^{N-1} (I_{ix} I_{(i+1)x} + I_{iy} I_{(i+1)y}),
\end{eqnarray}
where $\omega_0$ is the Larmour frequency in the external magnetic field, $D$ is the spin-spin coupling constant between the nearest neighbors and $I_{i\alpha}$ ($i=1,\dots,N$, $\alpha=x,y,z$) is the $i$th spin projection on the $\alpha$-axis. 
We chose the initial state of this chain with the single polarized $j$th node ($1\le j \le N$ ) at arbitrary temperature, i.e.
\begin{eqnarray}
\rho_0=\frac{e^{\beta I_{jz}}}{Z}=
\frac{1}{2^N}\left(1+2 I_{jz}\tanh \frac{\beta}{2}\right),\;\;Z={\mbox{Tr}} (e^{\beta I_{jz}}) = 2^N \cosh\frac{\beta}{2},
\end{eqnarray}
where $\beta=\frac{\hbar \omega_0}{kT}$, $\hbar$ is the Plank constant, $k$ is the Boltzmann constant, and $T$ is the temperature of the system. 
The evolution of the density matrix is described by the Liouville equation $\frac{d\rho}{dt}=-i[H,\rho]$ for the density matrix $\rho$. The solution to this equation reads:
\begin{eqnarray}\label{rho_t}
\rho(t)= e^{-i t H} \rho_0 e^{i t H}= 
\frac{1}{2^N} e^{-iH t} (1+2 I_{jz} \tanh\frac{\beta}{2}) e^{iH t}.
\end{eqnarray}
To solve the Liouville equation, we diagonalize the 
 Hamiltonian (\ref{XY})  using  
 the Jordan-Wigner transformation method \cite{JW}
\begin{eqnarray}
H=\sum_{k} \varepsilon_k \beta_k^+\beta_k -\frac{1}{2} N \omega_0,\;\;
\varepsilon_k = D \cos(k) +\omega_0,
\end{eqnarray}
where the fermion operators $\beta_j$ are defined in terms of   other fermion operators $c_j$  by means of the Fourier transformation
\begin{eqnarray}
\beta_k = \sum_{j=1}^N g_k(j) c_j,
\end{eqnarray}
and the fermion operators  $c_j$ are defined as \cite{JW}
\begin{eqnarray}
c_j=(-2)^{j-1} I_{1z}I_{2z}\dots I_{z(j-1)} I^-_j.
\end{eqnarray}
Here
\begin{eqnarray}
g_k(j)=\left(
\frac{2}{N+1}
\right)^{1/2} \sin(k j),\;\;\;\displaystyle k=\frac{\pi n}{N+1}, \;\;\;n=1,2,\dots,N.
\end{eqnarray}
It may be readily shown that the projection operators $I_{jz}$ can be  expressed in terms of the fermion operators $c_j$ as
\begin{eqnarray}
I_{jz} = c^+_j c_j -\frac{1}{2},\;\;\forall \; j.
\end{eqnarray}
Then the density matrix (\ref{rho_t})
can  be transformed to the following form \cite{FBE}
\begin{eqnarray}\label{rhot}
\rho(t)=\frac{1-\tanh\frac{\beta}{2}}{2^N} + \frac{\tanh\frac{\beta}{2}}{2^{N-1}} \sum_{k,k'}
e^{-i t(\varepsilon_k-\varepsilon_{k'})} g_k(j) g_{k'}(j) \beta^+_k\beta_{k'}.
\end{eqnarray}
where we use the  identity 
\begin{eqnarray}
e^{-i \varphi \beta^+_k \beta_k} \beta^+_k e^{i \varphi \beta^+_k \beta_k} = e^{-i \varphi} \beta^+_k,\;\;
\forall \;\varphi.
\end{eqnarray}
It is interesting to note that the quantity $\frac{\langle I_{pz}\rangle (t)}{\langle I_{jz}\rangle (0)}$ for the $p$th node does not depend on $\beta$ and coincides with the result of ref.\cite{FBE} obtained for the high temperatures only ($\langle a \rangle \equiv{\mbox{Tr}}\{\rho a\}$ for any operator $a$):
\begin{eqnarray}
\frac{\langle I_{pz}\rangle (t)}{\langle I_{jz}\rangle (0)}=\frac{4}{(N+1)^2} 
\left|
\sum_ke^{-i\varepsilon_k t} \sin(k j) \sin ( kp)
\right|^2.
\end{eqnarray}

We will study the quantum discord $Q_{nm}$ and the concurrence $C_{nm}$ (as a measure of entanglement) 
between any two particles. Since $Q_{nm}=Q_{mn}$ and $C_{nm}=C_{mn}$, hereafter we take $m>n$ without loss of generality, $n=1,\dots,N$.  As was mentioned in the Introduction,  we consider the open  spin-1/2 chain using its density matrix representations in three different bases:
(i) the basis of eigenvectors of the fermion  operators $\beta_j$ (we refer to this  representation of the density matrix as the $\beta$-representation), (ii) the basis of eigenvectors of the Fourier transformed fermion operators   $c_j$ (the $c$-representation of the density matrix), and (iii) the basis of  eigenvectors of the operators $I_{jz}$ ($j=1,\dots,N$). Of course, considering the correlations between the $n$th and the $m$th nodes  in all these cases we consider the quantum correlations between physically different particles. Thus, using either the $\beta$- or $c$-representation we consider the correlations  either between the $n$th and $m$th $\beta$-fermions 
  or between the $n$th and $m$th $c$-fermions, while using the basis of  eigenvectors of the operators $I_{jz}$, we elaborate quantum correlations between the natural  spin-1/2 particles. This is the reason of qualitative and quantitative difference  among the
discords and entanglements 
found in all these cases. 

First step in calculation of either discord or entanglement between the  $n$th and $m$th nodes  is the construction of the 
 reduced density matrix with respect to all nodes except for the $n$th and $m$th ones.
Below, we calculate the reduced density matrices  in three different bases mentioned above. 
In all cases we use notations 
\begin{eqnarray}\label{basis}
|nm\rangle=\{|00\rangle,|01\rangle,|10\rangle, |11\rangle\}
\end{eqnarray}
for the basis vectors,  where $n$ in the vector  $|n\rangle$  means  the different filling numbers in  the fermion bases or the excited spin ($n=1$) and the ground state spin ($n=0$) in the  basis of  eigenvectors of operators $I_{jz}$.   

\subsection{ The reduced density matrix in the basis of eigenvectors of the fermion operators $\beta_j$}
\label{Section:bas1}
We use the superscript "$\beta$" to specify the $\beta$-representations  of  quantum operators.
Let us  reduce  density matrix (\ref{rhot}) with respect to all fermions except for the $n$th and $m$th ones  obtaining:
\begin{eqnarray}
\rho_{nm}^\beta = \frac{1}{4} - \frac{\tanh\frac{\beta}{2} }{4} (g_n^2(j) +g_m^2(j)) +
\frac{\tanh\frac{\beta}{2} }{2} \sum_{k,k'=n,m} e^{-i t(\varepsilon_k-\varepsilon_k')} g_k(j) g_{k'}(j) \beta^+_k \beta_{k'}.
 \end{eqnarray}
Using the basis (\ref{basis}) we find
\begin{eqnarray}
\label{rhonm_matr}
\rho_{nm}^\beta= \left(
\begin{array}{cccc}
J^\beta_{00}  + J^\beta_{mm}+ J^\beta_{nn}&0&0&0\cr
0&J^\beta_{00}  + J^\beta_{mm}&J^\beta_{mn}&0\cr
0&J^\beta_{nm} &J^\beta_{00}  + J^\beta_{nn}&0\cr
0&0&0&J^\beta_{00}
\end{array}
\right),
\end{eqnarray}
where 
\begin{eqnarray}\label{J_bet}
&&
J^\beta_{00}= \frac{1}{4} - \frac{\tanh\frac{\beta}{2} }{4} (g_n^2(j) +g_m^2(j)) ,
\\\nonumber
&&
J^\beta_{nm}=\frac{\tanh\frac{\beta}{2} }{2}  e^{-i t(\varepsilon_n-\varepsilon_m)} g_n(j) g_{m}(j)
\end{eqnarray}
It is obvious that
\begin{eqnarray}\label{J_nn_beta}
J^\beta_{nn}=\frac{\tanh\frac{\beta}{2} }{2}  g_n^2(j),
\end{eqnarray}
which does not depend on the time $t$.

\subsection{The reduced density matrix in the basis of eigenvectors of the fermion operators $c_j$}
\label{Section:bas2}

In this case we write the density matrix (\ref{rhot}) as
\begin{eqnarray}\label{rhot_c}
\rho(t)=\frac{1-\tanh\frac{\beta}{2}}{2^N} + \frac{\tanh\frac{\beta}{2}}{2^{N-1}} \sum_{k,k',l,l'}
e^{-i t(\varepsilon_k-\varepsilon_{k'})} g_k(j) g_{k'}(j) g_k(l) g_{k'}(l') c^+_l c_{l'}.
\end{eqnarray}
 We use the superscript "$c$" to specify the $c$-representations of  quantum operators.  Reducing the density matrix $\rho$  with respect to all fermions except for the $n$th and $m$th ones, we  obtain 
\begin{eqnarray}
&&
\rho_{nm}^c = \frac{1}{4} - \frac{\tanh\frac{\beta}{2} }{4} \sum_{k,k'}
\sum_{l=n,m}e^{-i t(\varepsilon_k-\varepsilon_{k'})}g_k(j) g_{k'}(j) g_k(l)g_{k'}(l)+\\\nonumber
&&
\frac{\tanh\frac{\beta}{2} }{2} \sum_{k,k'} \sum_{l,l'=n,m} 
e^{-i t(\varepsilon_k-\varepsilon_k')} g_k(j) g_{k'}(j)  g_k(l) g_{k'}(l')c^+_l c_{l'}.
\end{eqnarray}
The matrix form of the reduced density matrix  $\rho^c_{nm}$ in the basis (\ref{basis}) coincides with  eq.(\ref{rhonm_matr}) up to the replacements 
$J^\beta_{nm}\to J^c_{nm}$
with
\begin{eqnarray}\label{J_c}
J^c_{00}&=&\frac{1}{4} - \frac{\tanh\frac{\beta}{2} }{4} \sum_{k,k'}
\sum_{l=n,m}e^{-i t(\varepsilon_k-\varepsilon_{k'})} g_k(j) g_{k'}(j) g_k(l)g_{k'}(l)=\\\nonumber
&& \frac{1}{4} -\frac{1}{2}(J^c_{nn}+J^c_{mm}),\\\nonumber
J^c_{nm}&=&\frac{\tanh\frac{\beta}{2} }{2} \sum_{k,k'} 
e^{-i t(\varepsilon_k-\varepsilon_k')} g_k(j) g_{k'}(j)  g_k(n) g_{k'}(m).
\end{eqnarray}
It is obvious that
\begin{eqnarray}\label{J_nn_c}
J^c_{nn}(t)=\frac{\tanh\frac{\beta}{2} }{2} \left |\sum_{k} 
e^{-i t\varepsilon_k} g_k(j)  g_k(n)\right |^2,\;\;\;J^c_{nn}(0)=\frac{\tanh\frac{\beta}{2} }{2}\delta_{jn}
,
\end{eqnarray}
where $\delta_{jn}$ is the Kronecker symbol.
Eq.(\ref{J_nn_c}) means, in particular, that 
$J^c_{nn}\le \frac{1}{2}$.

\subsection{The reduced density matrix in the basis of  eigenvalues of operators $I_{jz}$, $j=1,\dots,N$}
\label{Section:bas3}
In this case we write the density matrix as 
\begin{eqnarray}
&&
\rho(t)=\frac{1-\tanh\frac{\beta}{2}}{2^N} + \frac{\tanh\frac{\beta}{2}}{2^{N-1}} \sum_{k,k',l,l'}
e^{-i t(\varepsilon_k-\varepsilon_{k'})} g_k(j) g_{k'}(j) g_k(l) g_{k'}(l') \times \\\nonumber
&&
(-2)^{l+l'-2} I_{1z}\dots I_{(l-1)z} I^+_l I_{1z}\dots I_{(l'-1)z} I^-_{l'}.
\end{eqnarray}
We use the superscript "$spin$" to specify the matrix representations of  quantum operators in the basis of eigenvectors of the  operators $I_{jz}$.
Reducing the density matrix $\rho$  with respect to all nodes except for the $n$th and $m$th ones, we obtain 
that the reduced density matrix is non-diagonal only if $m=n\pm 1$:
\begin{eqnarray}
&&
\rho^{spin}_{n(n+1)} = \frac{1}{4} -\frac{\tanh\frac{\beta}{2}}{4} 
\sum_{k,k'}\sum_{l=n,n+1}e^{-it (\varepsilon_k-\varepsilon_{k'})} g_k(j) g_{k'}(j) g_k(l) g_{k'}(l) +
\\\nonumber
&&
\frac{\tanh\frac{\beta}{2}}{2}  \sum_{k,k'}e^{-it (\varepsilon_k-\varepsilon_{k'})} g_k(j)g_{k'}(j)
\Big(
g_k(n) g_{k'}(n+1) I_n^+ I_{n+1}^- + g_k(n+1) g_{k'}(n) I_{n+1}^+
I_{n}^-+\\\nonumber
&&
g_k(n) g_{k'}(n) I_n^+ I_{n}^-+g_k(n+1) g_{k'}(n+1) I_{n+1}^+ I_{n+1}^-
\Big) 
\end{eqnarray}
Otherwise ($m\neq n\pm 1$) the density matrix is diagonal:
\begin{eqnarray}
&&
\rho^{spin}_{nm} = \frac{1}{4} -\frac{\tanh\frac{\beta}{2}}{4} 
\sum_{k,k'}\sum_{l=n,m}e^{-it (\varepsilon_k-\varepsilon_{k'})} g_k(j) g_{k'}(j) g_k(l) g_{k'}(l) +
\\\nonumber
&&
\frac{\tanh\frac{\beta}{2}}{2}  \sum_{k,k'}e^{-it (\varepsilon_k-\varepsilon_{k'})} g_k(j)g_{k'}(j)
\Big(
g_k(n) g_{k'}(n) I_n^+ I_{n}^-+g_k(m) g_{k'}(m) I_{m}^+ I_{m}^-
\Big) ,
\end{eqnarray}
so that the discord and entanglement are both zero.
The matrix form of the reduced density matrix  $\rho^{spin}_{n(n+1)}$ in the basis (\ref{basis}) coincides with  eq.(\ref{rhonm_matr}) 
up to the replacements 
$J^\beta_{nm}\to J^{spin}_{nm}$
with
\begin{eqnarray}\label{J_Iz}
J^{spin}_{00}=J^{c}_{00},\;\;J^{spin}_{nn}=J^{c}_{nn},\;\;J^{spin}_{(n+1)(n+1)}=J^{c}_{(n+1)(n+1)},\;\;J^{spin}_{n(n+1)}=J^{c}_{n(n+1)}.
\end{eqnarray}


\section{General formulas for discord and concurrence}
\label{Section:discord}

Since the reduced bi-particle density matrices in all three cases have the same matrix form (\ref{rhonm_matr}), we derive the formulas for both the concurrence (as a measure of entanglement) and   the discord for the density matrix (\ref{rhonm_matr}) omitting the  superscript $\beta$.
We will use the relation
\begin{eqnarray}\label{Jnm}
|J_{mn}|^2=J_{nn} J_{mm}
\end{eqnarray}
which is valid in all three cases considered in  Secs.\ref{Section:bas1}-\ref{Section:bas3}.

\subsection{Concurrence}
We characterize the entanglement  by the Wootters criterion in terms of the concurrence \cite{HW,Wootters}.
According to \cite{HW,Wootters}, 
one needs to construct the spin-flip density matrix
\begin{eqnarray}\label{trho2}
\tilde\rho_{(nm)}(\tau) =(\sigma_{y}\otimes \sigma_y) (\rho^{(nm)})^*(\tau) (\sigma_y\otimes \sigma_y),
\end{eqnarray}
where the asterisk denotes the complex conjugation in the  basis (\ref{basis})
and the Pauli matrix
$\sigma_y= 2 I_y$. The concurrence for the density matrix $\rho_{(nm)}(\tau)$ is equal to 
\begin{eqnarray}\label{C}
C=\max(0,2\lambda -\lambda_1-\lambda_2-\lambda_3-\lambda_4),\;\;\lambda=\max(\lambda_1,\lambda_2,\lambda_3,\lambda_4),
\end{eqnarray}
where $\lambda_1$, $\lambda_2$, $\lambda_3$ and $\lambda_4$ are the square roots of the eigenvalues of the matrix product 
$\rho_{(nm)}(\tau) \tilde \rho_{(nm)}(\tau)$.
For the density matrix $\rho$ given by eq.(\ref{rhonm_matr}) we have
\begin{eqnarray}\label{lambdas}
&&
\lambda=\lambda_1=\frac{1}{4}\sqrt{1-4(J_{mm} - J_{nn})^2} +\sqrt{J_{mm} J_{nn}},\\\nonumber
&&
\lambda_2=\frac{1}{4}\sqrt{1-4(J_{mm} - J_{nn})^2} -\sqrt{J_{mm} J_{nn}},\\\nonumber
&&
\lambda_{3}=\lambda_4=\frac{1}{4} \sqrt{1-4(J_{mm}+J_{nn})^2}.
\end{eqnarray}
Substituting eqs.(\ref{lambdas}) into eq.(\ref{C}) we obtain
\begin{eqnarray}\label{C2}
C_{nm}=\max\left(0,2\sqrt{J_{mm} J_{nn}} -\frac{1}{2}\sqrt{1-4(J_{mm} + J_{nn})^2} \right).
\end{eqnarray}

\subsection{Discord}
As far as  matrix 
(\ref{rhonm_matr}) is a particular case of the X-matrix,
we will use the results of ref. \cite{ARA}, where the discord for the X-matrix has been studied.
We remind the basic formulas for calculation of the discord $Q^{(B)}$ between two particles, which are called the subsystems $A$ and $B$ \cite{ARA}. Recall, that the superscript $B$ means that the projective measurements are performed over the subsystem $B$.
The discord is introduced as a difference between the total mutual information $I(\rho)$ encoded into the system $AB$ and its classical part ${\cal{C}}^B$,
\begin{eqnarray}\label{Q}
Q^{(B)}= I(\rho) - {\cal{C}}^{(B)},
\end{eqnarray}
The total mutual information $I$  reads
\begin{eqnarray}\label{I}
I(\rho)=S(\rho^{(A)}) + S(\rho^{(B)}) +\sum_{j=0}^3 \lambda_j\log_2\lambda_j.
\end{eqnarray}
Here $\rho^{(A)}$ and  $\rho^{(B)}$ of the  reduced density matrices of the subsystems $A$ and $B$,    $S(\rho^{(A)}) $ and  $S(\rho^{(B)})$ are the  von Neumann entropies of the  subsystems $A$ and $B$,
\begin{eqnarray} \label{SA}
&&
S(\rho^{(A)}) = -\frac{1}{2}\left(\log_2(\frac{1}{4}-J_{mm}^2) + 
2 J_{mm} \log_2\frac{1+2 J_{mm}}{1-2J_{mm}}\right),\\\nonumber
&&
S(\rho^{(B)}) = -\frac{1}{2}\left(\log_2(\frac{1}{4}-J_{nn}^2) + 
2 J_{nn} \log_2\frac{1+2 J_{nn}}{1-2J_{nn}}\right),
\end{eqnarray}
 $\lambda_j$ are the eigenvalues of  the density matrix (\ref{rhonm_matr}),
\begin{eqnarray}\label{lam}
\lambda_{0,1}=\frac{1}{4}(1- 2(J_{mm}+J_{nn})),\;\;\lambda_{2,3}=\frac{1}{4}(1 + 2(J_{mm}+J_{nn})).
\end{eqnarray}
As far as all eigenvalues of any density matrix  must be non-negative, eqs.(\ref{lam}) mean
\begin{eqnarray}
\label{JJnnmm}
J_{mm}+J_{nn}\le 1/2.
\end{eqnarray}

The classical counterpart ${\cal{C}}^{(B)}$ reads \cite{ARA}
\begin{eqnarray}\label{cl_min}
{\cal{C}}^{(B)} = S(\rho^{(A)}) - \min_{\eta=\{0,1\}} (p_0 S_0  + p_1 S_1).
\end{eqnarray}
Here the conditional entropies $S_i$ ($i=0,1$) are
\begin{eqnarray}
&&
S_i=-\frac{1-\theta^{(i)}}{2}\log_2 \frac{1-\theta^{(i)}}{2}-
\frac{1+\theta^{(i)}}{2}\log_2 \frac{1+\theta^{(i)}}{2},\\\nonumber
&&
\theta^{(0)}=\frac{2\sqrt{J_{mm}(J_{mm}-(\eta^2-1) J_{nn})}}{1+2\eta J_{nn}},\;\;
\theta^{(1)}=\frac{2\sqrt{J_{mm}(J_{mm}-(\eta^2-1) J_{nn})}}{1-2\eta J_{nn}}
\end{eqnarray}
and the populations $p_i$, $i=0,1$ are
\begin{eqnarray}
&&
p_0=\frac{1}{2}(1+2\eta J_{nn}) ,\;\;p_1=\frac{1}{2}(1-2\eta J_{nn}),
\end{eqnarray}
where we introduce the parameter $\eta$ instead of $k$ used in ref.\cite{ARA}, $k=\frac{1+\eta}{2}$.
It might be readily demonstrated that the minimum in eq.(\ref{cl_min}) corresponds to $\eta=0$. For this purpose we show that the  derivative of the function
\begin{eqnarray}
f(\eta)=
p_0 S_0  + p_1 S_1
\end{eqnarray}
with respect to the parameter $\eta$  is positive over the interval $0\le  \eta \le 1$.
In fact, this derivative reads after some transformations:
\begin{eqnarray}\label{der2}
&&
f'(\eta)=(\alpha+\beta)\log_2\frac{1-2\eta J_{nn} + 2 d }{1+2\eta J_{nn} - 2 d} +\\\nonumber
&&
(\beta-\alpha)\log_2\frac{1+2\eta J_{nn} + 2 d }{1-2\eta J_{nn} - 2 d} +
2 \alpha \log_2\frac{1+2\eta J_{nn}}{1-2\eta J_{nn}} ,\\\nonumber
&&\alpha=\frac{J_{nn}}{2} ,\;\;\beta=\frac{\eta J_{mm} J_{nn}}{2d},\;\;d= \sqrt{J_{mm}(J_{mm} -
(\eta^2-1) J_{nn})}. 
\end{eqnarray}
Collecting terms with $\alpha$ and $\beta$ we obtain:
\begin{eqnarray}\label{der3}
f'(\eta)=\alpha\log_2\frac{  4 d^2(1+2\eta J_{nn})^2- (1-4\eta^2 J_{nn}^2)^2 }{  4 d^2(1-2\eta J_{nn})^2- (1-4\eta^2 J_{nn}^2)^2 }
+
\beta \log_2\frac{(1+2d)^2 -  4\eta^2 J_{nn}^2}{(1-2d)^2 -  4\eta^2 J_{nn}^2}.
\end{eqnarray}
Both terms in the RHS  of eq.(\ref{der3}) are nonnegative. Thus we conclude that $f'(\eta)\ge 0$, i.e. $f(\eta)$ is increasing function over the interval $0\le \eta\le 1$. Consequently, the   function $f(\eta)$ takes the minimal value  at the boundary point $\eta=0$.

As a result, the expression for the classical correlations (\ref{cl_min}) reads:
\begin{eqnarray}\label{cl}
{\cal{C}}^{(B)}&=& 
\frac{1}{2}\log_2\frac{1-4J_{mm}(J_{mm}+J_{nn})}{1-4J_{mm}^2}  +
J_{mm} \log_2\frac{1-2 J_{mm}}{1+2 J_{mm}} + \\\nonumber
&&
\sqrt{J_{mm}(J_{mm} +J_{nn})} \log_2\frac{1+2\sqrt{J_{mm}(J_{mm}+J_{nn})}}{1-2\sqrt{J_{mm}(J_{mm}+J_{nn})}}
\end{eqnarray}
Now the discord $Q^{(B)}$ may be calculated by  formula (\ref{Q}) using eqs.(\ref{I}-\ref{lam}) and (\ref{cl}):
\begin{eqnarray}\label{finaldiscord}
Q^{(B)}&=&-\frac{1}{2}\Big(
(1-2 J_{nn}) \log_2 (1-2 J_{nn})+(1+2 J_{nn}) \log_2 (1+2 J_{nn})-\\\nonumber
&&
(1-2 J_{mm} - 2 J_{nn}) \log_2(1-2 J_{mm} - 2 J_{nn})-\\\nonumber
&&
(1+2 J_{mm} + 2 J_{nn}) \log_2(1+2 J_{mm} + 2 J_{nn})+\\\nonumber
&&
(1-2 \sqrt{J_{mm}(J_{mm}+J_{nn})})\log_2(1-2 \sqrt{J_{mm}(J_{mm}+J_{nn})})
+\\\nonumber
&&
(1+2 \sqrt{J_{mm}(J_{mm}+J_{nn})})\log_2(1+2 \sqrt{J_{mm}(J_{mm}+J_{nn})})\Big).
\end{eqnarray}
Note that the discord $Q^{(A)}$ (obtained by means of the Neumann type measurements over the subsystem $A$) differs from the discord $Q^{(B)}$ by the replacement $n \leftrightarrow m$. As far as $Q^{(A)}\neq Q^{(B)}$ in general \cite{HV,OZ}, 
we define the discord $Q_{nm}$ as follows \cite{FZ}:
\begin{eqnarray}\label{def_discord}
Q_{nm}=\min(Q^{(A)}_{nm},Q^{(B)}_{nm}).
\end{eqnarray}

\section{Discord and concurrence in different  density matrix representations of the open spin chain with $N\gtrsim 10$ nodes }
\label{Section:comparison}

The problem of description of those quantum correlations which are responsible for advantages of quantum computations is not resolved yet. Therefore it is reasonable to characterize different types of quantum correlations in a quantum system, i.e. not only those which  correspond to the  real particles, but also correlations between the ''virtual particles''. 
This was the basic motivation for introduction of the  unitary invariant discord   in ref.\cite{Z_a}   as a way to count all quantum correlations between all possible ''virtual particles''. The geometric measure $Q^G$ of the unitary invariant discord is  representable in terms of eigenvalues $\lambda_i$ of the considered density matrix,
\begin{eqnarray}
Q^G =\frac{2^N\sum_{i=1}^{2^N} \lambda_i^2 -1 }{2^N-1}, 
\end{eqnarray}
so that it achieves the maximal value 1 for a pure state and is zero for the matrix with all equal eigenvalues.  In the case of bi-particle density matrix (\ref{rhonm_matr}), $N=2$, with eigenvalues (\ref{lam}) we  obtain 
\begin{eqnarray}\label{qg}
Q^G_{nm}=\frac{4}{3} (J_{mm}+J_{nn})^2
\le \frac{1}{3}.
\end{eqnarray}
The significant value of $Q^G$ suggests us to look for such basis where the quantum correlations are valuable and/or the evolution of quantum correlations is suitable for applications. 

In the present paper we investigate  the quantum correlations in a spin-1/2 chain of $N$ nodes    using three different matrix representations, i.e. we consider the same chain using 
(i) the
 $\beta$-representation introduced in  Sec.\ref{Section:bas1}, (ii)  the 
 $c$-representation introduced in Sec.\ref{Section:bas2}, and  (iii)  the representation in 
 the basis of eigenvectors of the operators $I_{jz}$, $j=1,\dots,N$, discussed in Sec.\ref{Section:bas3}. Below we study the quantum correlations in all three bases and demonstrate the significant qualitative and quantitative distinctions  among the
  discord and the concurrence in all three cases. It will be illustrated that the discord in the $\beta$- and $c$-representations of the density matrix reveals the new properties of the considered quantum system  in comparison with the usual  discord in the basis of eigenvectors of the operators $I_{jz}$.


\subsection{Discord and concurrence in a system of $\beta$-fermions}
\label{Section:beta_repr}
First,  we consider the $\beta$-representation of the density matrix (see Sec.\ref{Section:bas1}) and calculate the discord and the concurrence between the $n$th and $m$th fermions.  
Formula  (\ref{C}) for  the concurrence  reads in this case:
\begin{eqnarray}
C^\beta_{nm}=\max\Big( 0, \tanh\frac{\beta}{2} g_m(j) g_n(j)-
\frac{1}{2} \sqrt{1-\tanh^2\frac{\beta}{2}(g_m^2(j) + g_n^2(j))^2} \Big).
\end{eqnarray}
Simple estimation at $\beta\to\infty$ shows that the concurrence may be positive only for $N\le 4$, which is not interesting for us. In the case of longer chain, $C^\beta_{nm}=0$ for any nodes $n$ and $m$.

\begin{figure*}
   \epsfig{file=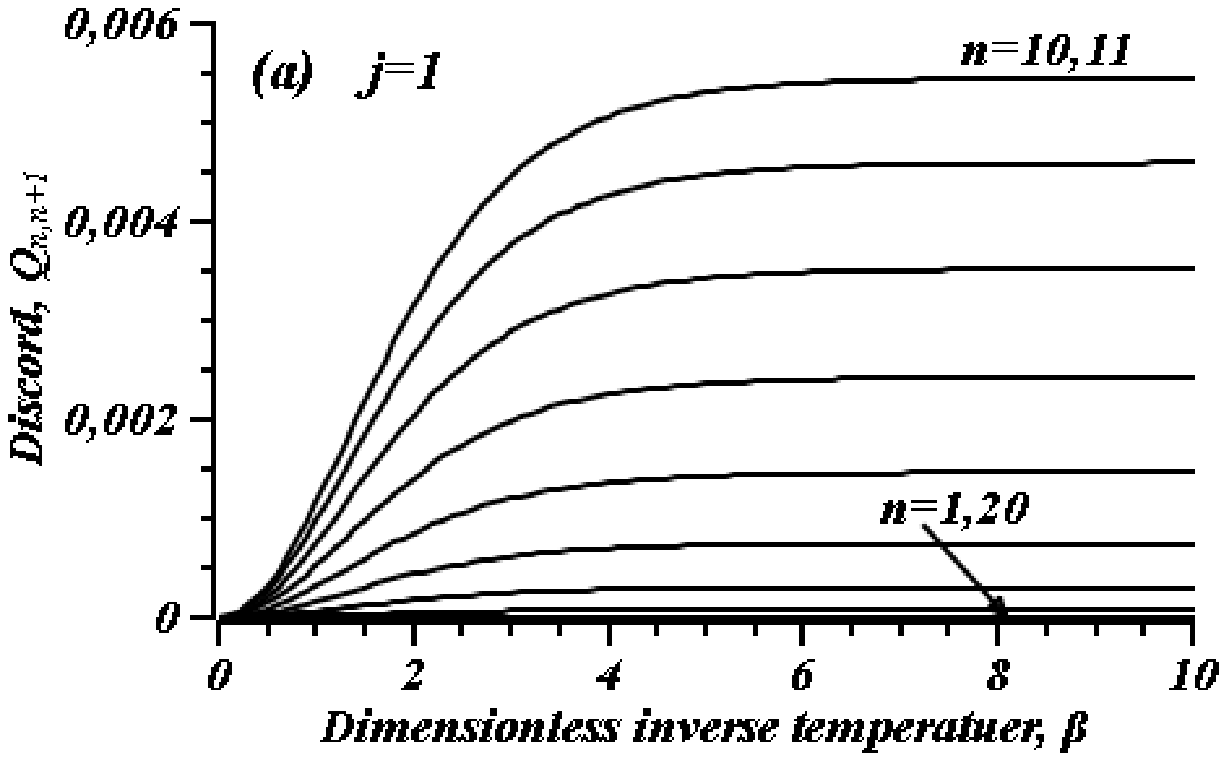, 
    scale=0.6
   ,angle=0
}
\epsfig{file=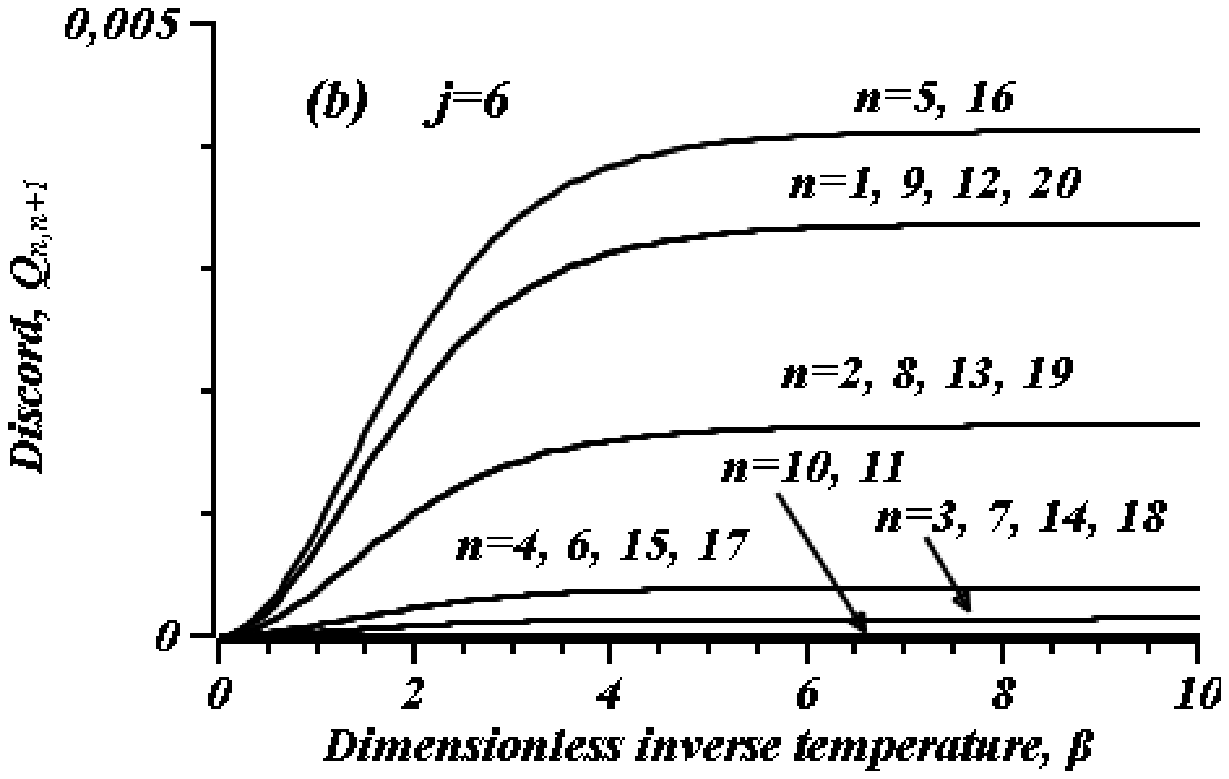
   , scale=0.6,angle=0
}
  \caption{ The discord $Q^\beta_{n,n+1}$ versus the dimensionless inverse temperature $\beta$ for the spin chain with $N=21$; $(a)$ the 1st node is initially polarized ($j=1$), the discord increases from the ends  to the center of the chain; $(b)$  the 6th node is initially polarized ($j=6$), the discord is a periodic function of the node $n$.}
  \label{Fig:bet_d} 
\end{figure*}

\begin{figure*}
   \epsfig{file=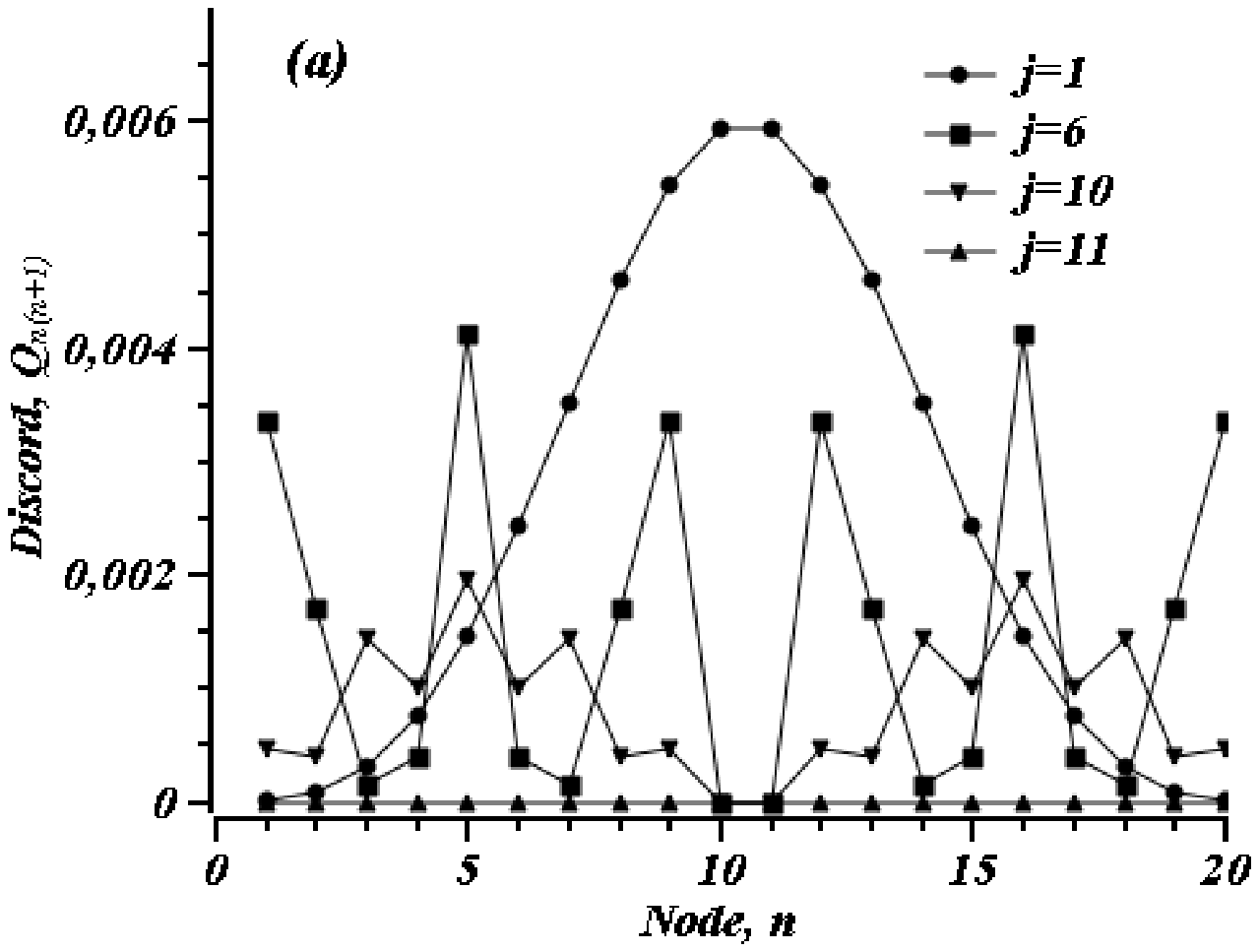
   , scale=0.6,angle=0
}
\epsfig{file=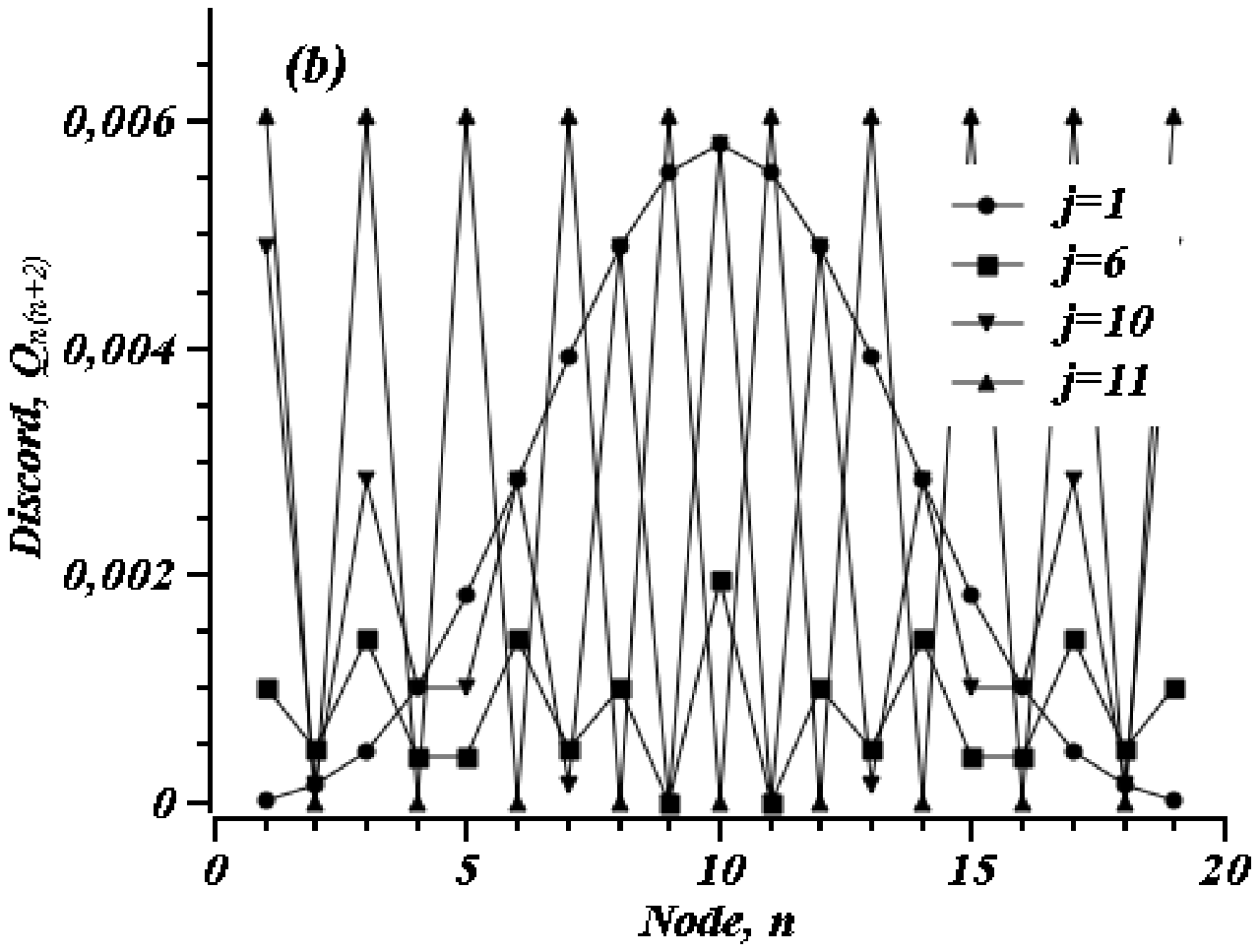
   , scale=0.6,angle=0
}\\
\epsfig{file=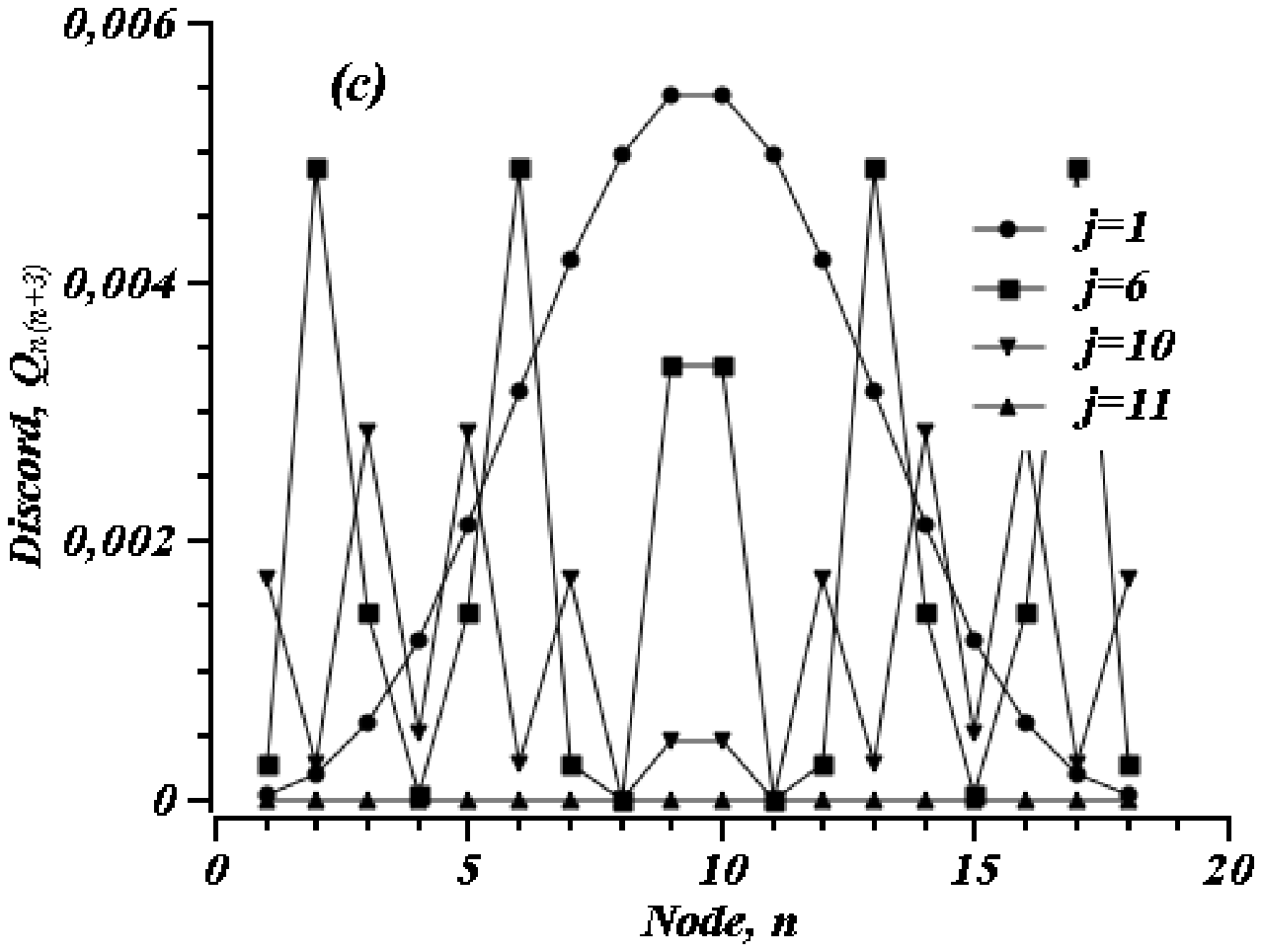
   , scale=0.6,angle=0
}
\epsfig{file=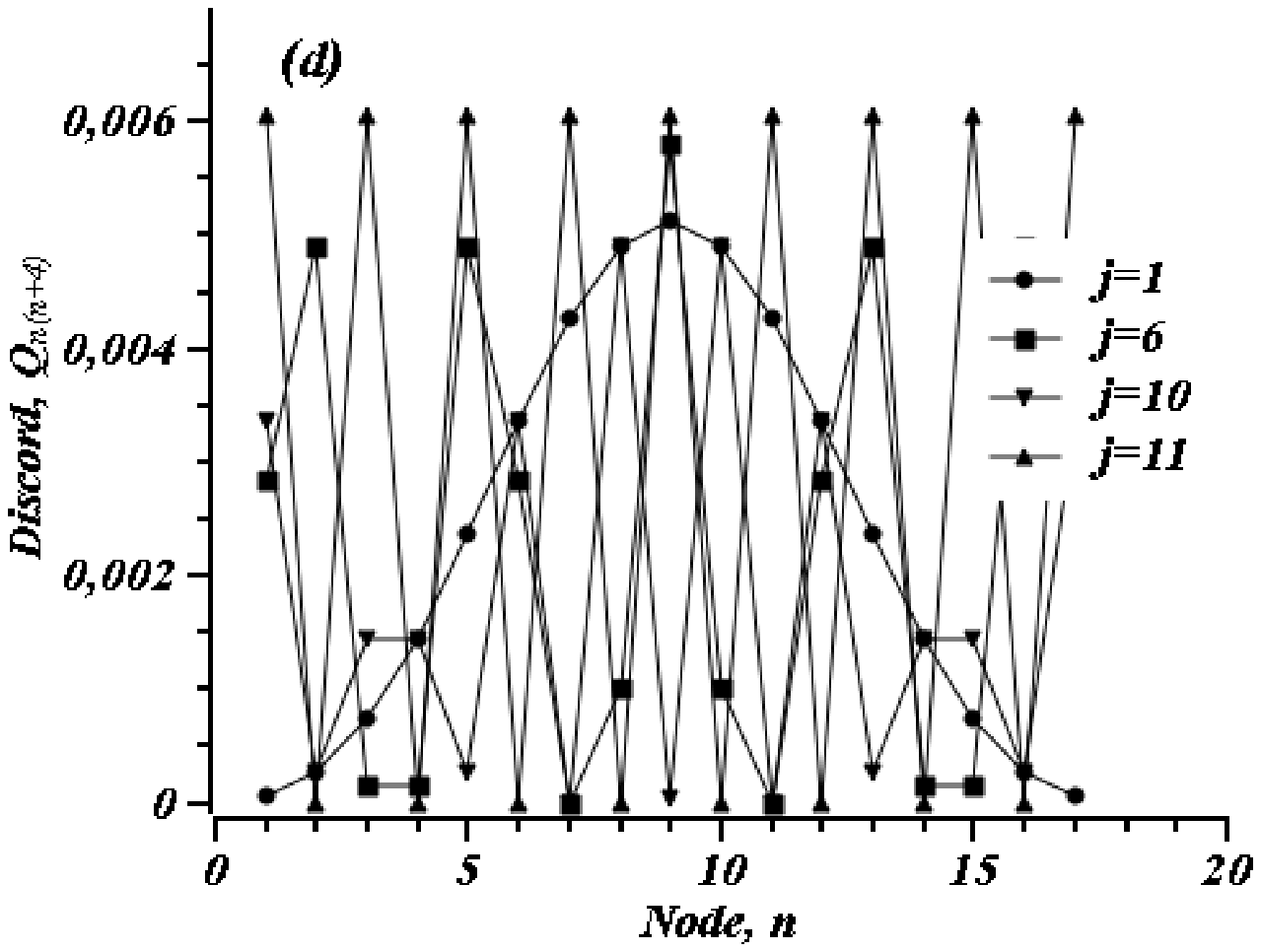
   , scale=0.6,angle=0
}
  \caption{ The discord $Q^\beta_{n,n+l}$, $l=1,2,3,4$, versus the node $n$ for the chain of  $N=21$ nodes  and 
  different initially polarized nodes $j=1,6,10,11$}
  \label{Fig:bet} 
\end{figure*}

The behavior of the discord is completely different. Using formulas 
(\ref{finaldiscord}) and (\ref{J_bet})  we obtain that the discord is non-zero in general and does not evolve in time. 
The discord $Q_{n,n+1}$ between the nearest neighbors versus $\beta=\frac{\hbar\omega_0}{kT}$ for $N=21$ is depicted in  Fig.\ref{Fig:bet_d}.
Emphasize that  the  discord is non-zero not only between the nearest  neighbors, but between the remote  nodes as well, which is not  valid in general, see, for instance, Sec.\ref{Section:spins}. 
The dependence of the discord $Q^\beta_{n,n+l}$, $l=1,2,3,4$,  on $n$ for $N=21$ and large values of $\beta$ ($\beta=10$) is shown in Fig.\ref{Fig:bet} for different initially polarized nodes $j=1,6,10,11$.
Fig.\ref{Fig:bet} reveals the interesting properties of discord.
\begin{enumerate}
\item
The discord possesses the following symmetry
 $Q^\beta_{n,n+k}=Q^\beta_{N-n,N-n+k}$, which is a consequence of the equality 
 \begin{eqnarray}
 g_{\frac{\pi n}{N+1}}(j)=g_{\frac{\pi (N+1-n)}{N+1}}(j)
 \end{eqnarray}
 and definition (\ref{def_discord}).
\item
If $N$ is odd and the middle node is initially polarized (i.e. $j=\frac{N+1}{2}$), then the discord $Q^\beta_{n,n+2k+1}$, $k=0,1,\dots$ is zero, while the discord $Q^\beta_{n,n+2k}$, $k=1,\dots$ is  zero only for even $n$, namely
\begin{eqnarray}
&&
Q^\beta_{n,n+2k}=\left\{\begin{array}{ll}
0,& n=2 i, \;\;i=1,2, \dots\cr
q, &n=2 i +1, \;\;i=0,1,\dots
\end{array}\right.
k=1,2\dots
\\\nonumber
&&
q=\frac{1}{2}\log_2\frac{\Big((N+1)^2-\left(4\tanh\frac{\beta}{2}\right)^2\Big) (N+1)^2}{
\left((N+1)^2-\left(2\tanh\frac{\beta}{2}\right)^2\right) \left((N+1)^2-2\left(2\tanh\frac{\beta}{2}\right)^2\right)
}+\\\nonumber 
&&
\frac{\tanh\frac{\beta}{2}}{N+1}\log_2\frac{
(N+1+4\tanh\frac{\beta}{2})^2(N+1-2\tanh\frac{\beta}{2})
}{(N+1-4\tanh\frac{\beta}{2})^2(N+1+2\tanh\frac{\beta}{2})}+\\\nonumber
&&\frac{\sqrt{2}\tanh\frac{\beta}{2}}{N+1}\log_2 
\frac{(N+1-2\sqrt{2}\tanh\frac{\beta}{2})}{(N+1+2\sqrt{2}\tanh\frac{\beta}{2})}.
\end{eqnarray}
Here   $q$ depends only on $\beta$ and $N$ and does not depend on the particular choice of odd $n$ and integer $k$.
The reason is that 
\begin{eqnarray}
\left|g_{k=\frac{\pi n}{N+1}}\left(\frac{N+1}{2}\right)\right|=\left(\frac{2}{N+1}\right)^{1/2} \left|\sin\frac{\pi n}{2}\right| =\left\{\begin{array}{ll}
0,& n=2 i ,\;\;i=1,2,\dots\cr
\left(\frac{2}{N+1}\right)^{1/2} ,& n=2 i+1,\;\;i=0,1,\dots.
\end{array}\right.
\end{eqnarray}
For the case $N=21$ and $\beta=10$ we obtain  $q\approx 0.0061$. 
The direct consequence of this property is an equality
\begin{eqnarray}\label{odd}
Q_{(2i_1+1)(2i_1+2l_1+1)}= Q_{(2i_2+1)(2i_2+2l_2+1)}, \;\;\forall i_1,i_2,l_1,l_2.
\end{eqnarray}
Eq.(\ref{odd}) means that we have the system of fermions (the odd nodes) with equal discords between any two of them. This system is a good candidate for the quantum register.
\item
If $j=1$, then the profile of the discord $Q^\beta_{n(n+l)}$, $\forall l$, is bell-shaped  with the maximum $Q^{max}_l$ in the node $[\frac{N+1-l}{2}]$ (here $[a]$ means the integer part of $a$) for discord $Q^\beta_{n(n+l)}$. If $l$ is even, 
then  we have $g_n(1)|_{n=[\frac{N+1-l}{2}]}=g_{n+l}(1)|_{n=[\frac{N+1-l}{2}]}=\sqrt{\frac{2}{N+1}} \cos\frac{\pi l}{2(N+1)}$ in eq.(\ref{J_nn_beta}). If $l$ is odd, 
then  we have $g_n(1)|_{n=[\frac{N+1-l}{2}]}=\sqrt{\frac{2}{N+1}} \cos\frac{\pi (l+1)}{2(N+1)}$,
$g_{n+l}(1)|_{n=[\frac{N+1-l}{2}]}=\sqrt{\frac{2}{N+1}} \cos\frac{\pi (l-1)}{2(N+1)}$ in eq.(\ref{J_nn_beta}).
 In our example ($\beta=10$, $N=21$) we have
$Q^{max}_1\approx 0.0059$, $Q^{max}_2\approx 0.0058$, $Q^{max}_3\approx 0.0055$, $Q^{max}_4\approx 0.0051$.
It is remarkable that $Q_{n(n+l)}\neq 0$ for all possible $n$ and $l$, i.e. all nodes are correlated. This is a principal advantage of 
 the case $j=1$ in comparison with  $j>1$, when the discord between some nodes is zero, see Fig.\ref{Fig:bet})   
\end{enumerate}
Finally remark, that the unitary invariant discord $Q^G_{n(n+l)}$ \cite{Z_a} as a function of the node $n$ for any fixed $l$ reproduces the shape of $Q^\beta_{n(n+l)}$ if $j=1$ and $11$, but $Q^G$ is bigger then $Q^\beta$. Thus, if $j=1$, then  $Q^G_{n(n+l)}$ is bell-shaped with the maximal values  $Q^G_{max}$ equal $0.0108$, $0.0106$, $0.0099$, $0.0093$ for  $l=1,2,3,4$ respectively;  if $j=11$ and  $l$ is odd, then the unitary invariant discord is the same between any neighbors, $Q^G_{n(n+l)}\approx 0.0028$; 
 if $j=11$ and  $l$ is even, then the unitary invariant discord  is saw-shaped with the amplitude $0.0110$. 

\subsection{Discord and concurrence in a system of $c$-fermions}
\label{Section:c_repr}
\begin{figure*}
\begin{minipage}{7cm}
   \epsfig{file=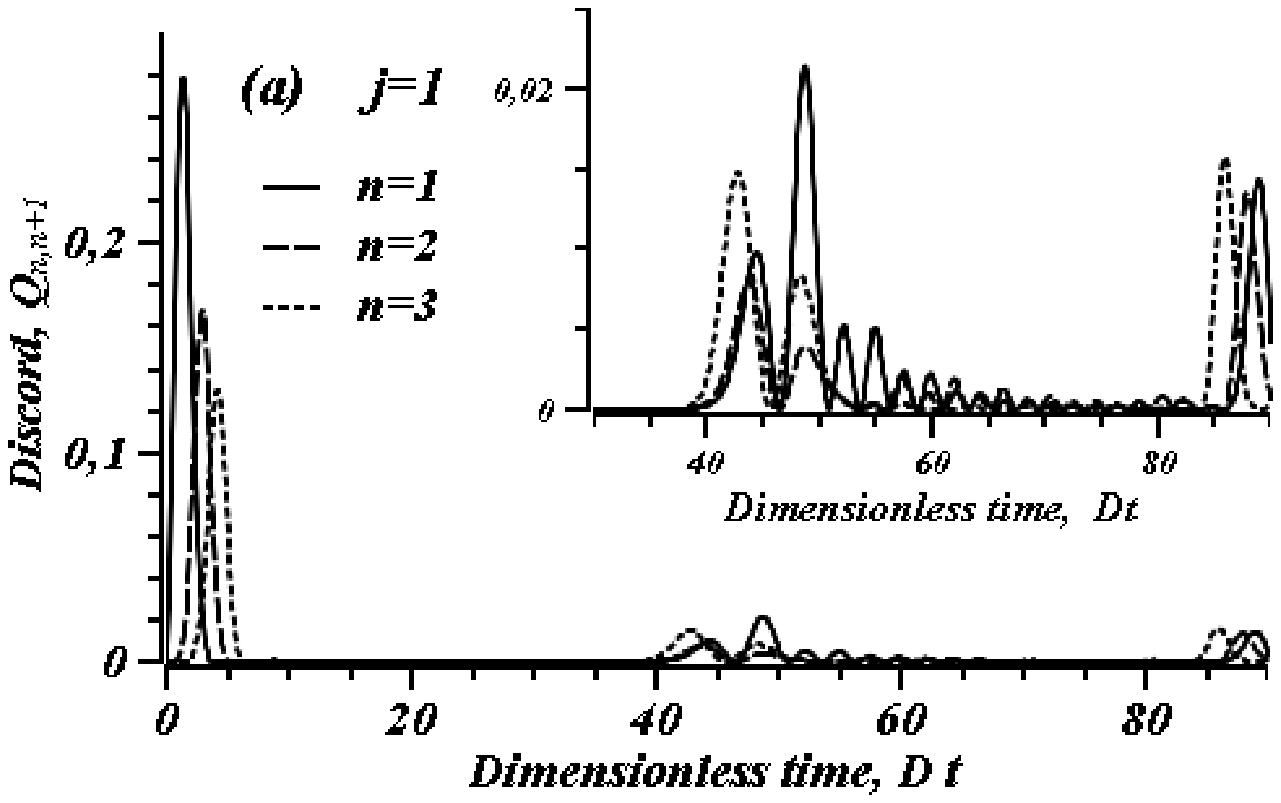
   , scale=0.6,angle=0
}\\
\epsfig{file=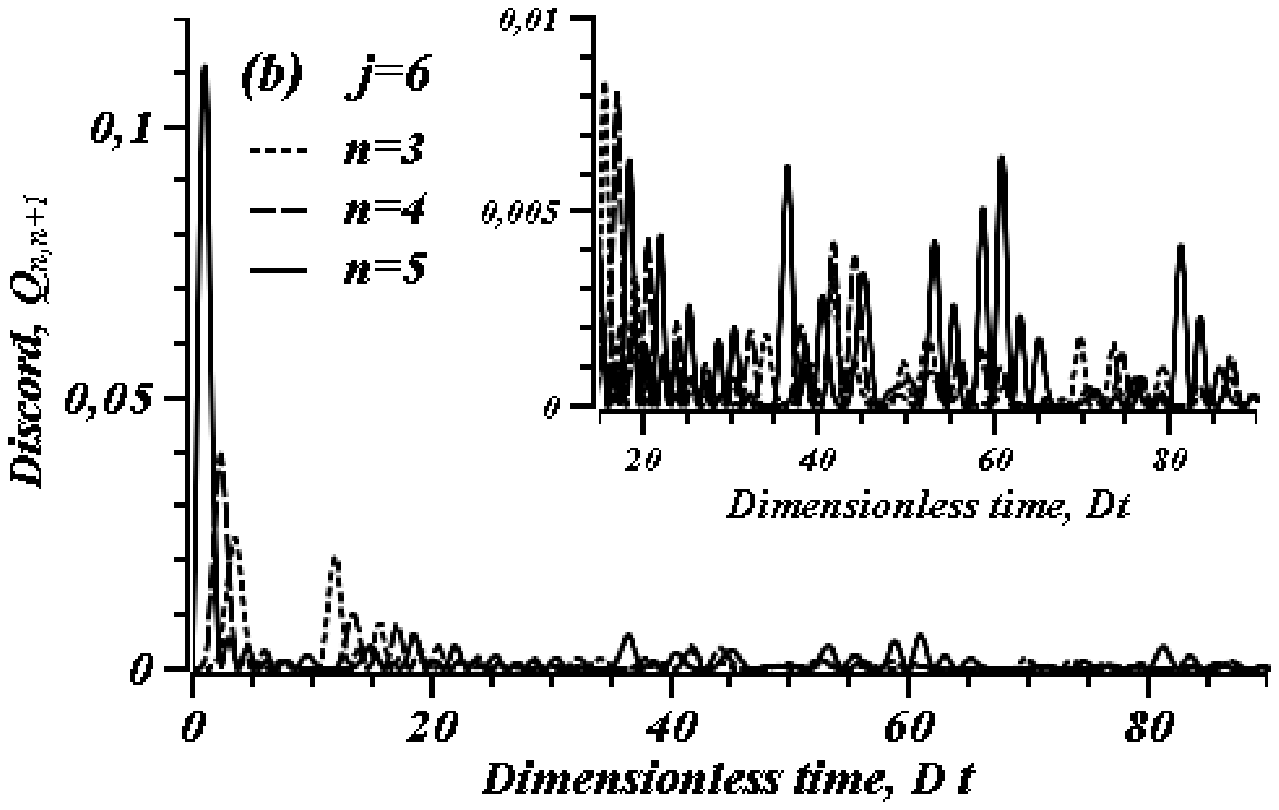
   , scale=0.6,angle=0
}\\
\epsfig{file=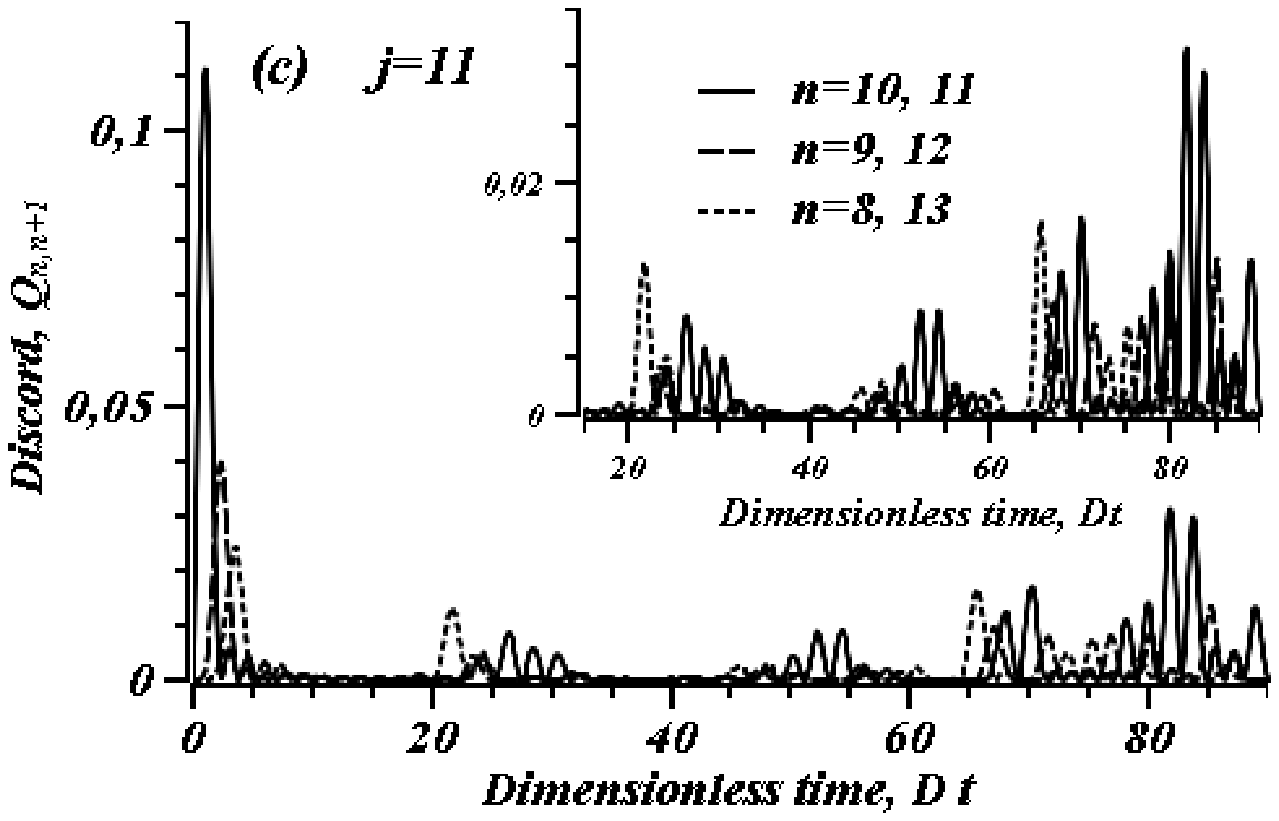
   , scale=0.6,angle=0
}
\end{minipage}\hspace{2cm}
\begin{minipage}{7cm}
 \epsfig{file=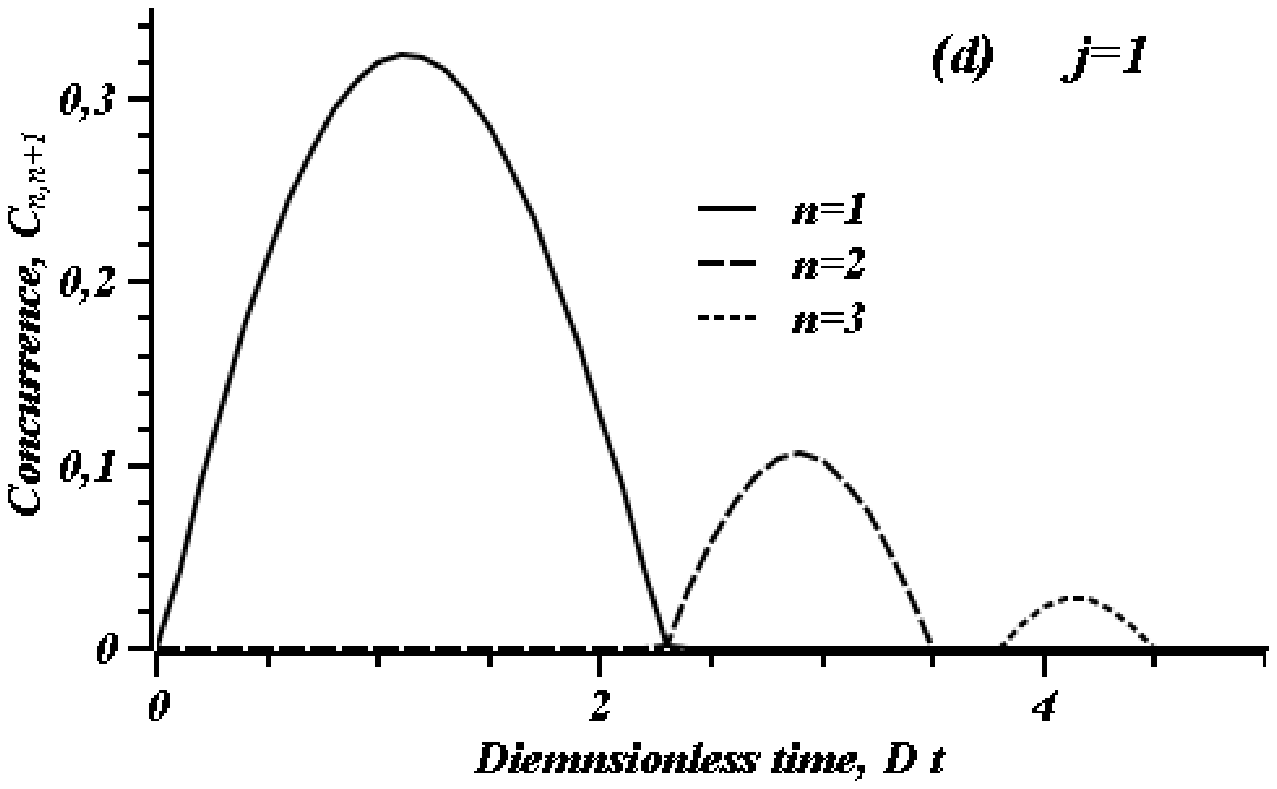
   , scale=0.6,angle=0
}\\
\epsfig{file=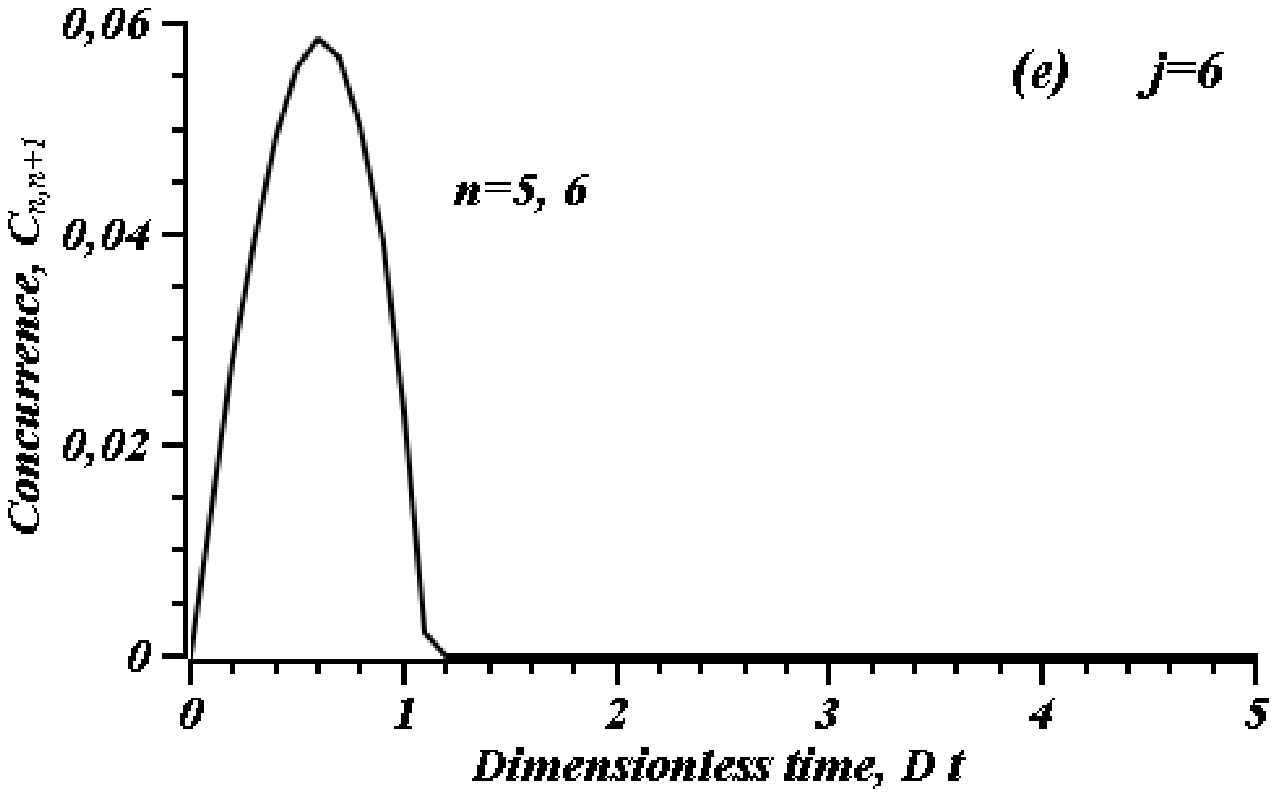
   , scale=0.6,angle=0
}\\
\epsfig{file=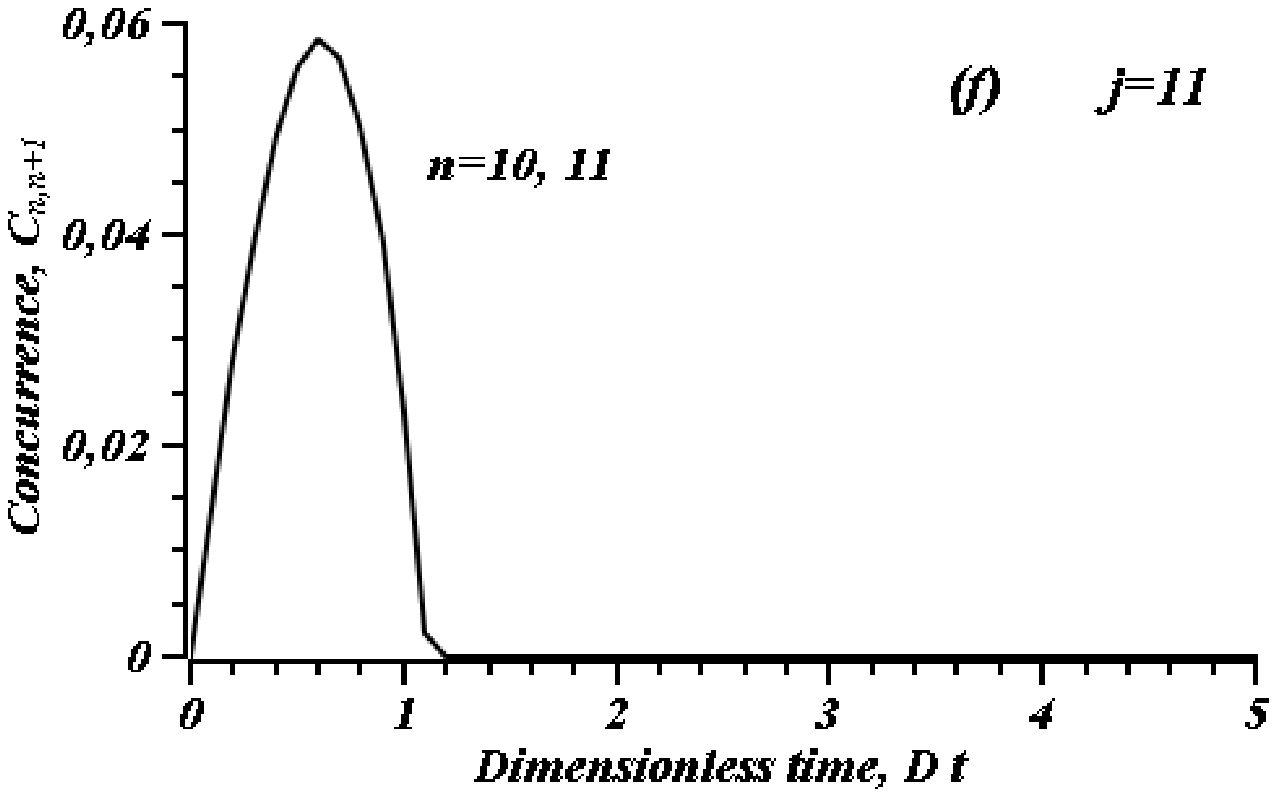
   , scale=0.6,angle=0
}\end{minipage}
  \caption{ The time evolution of the discord $Q^c_{n,n+1}$ $(a-c)$ and the concurrence $C^c_{n,n+1}$ $(d-f)$ 
  for $N=21$ and different initially polarized nodes $j=1,6,11$. The insets illustrate the ''echo'' in the evolution of quantum discord}
  \label{Fig:c_C} 
\end{figure*}

\begin{figure*}
   \epsfig{file=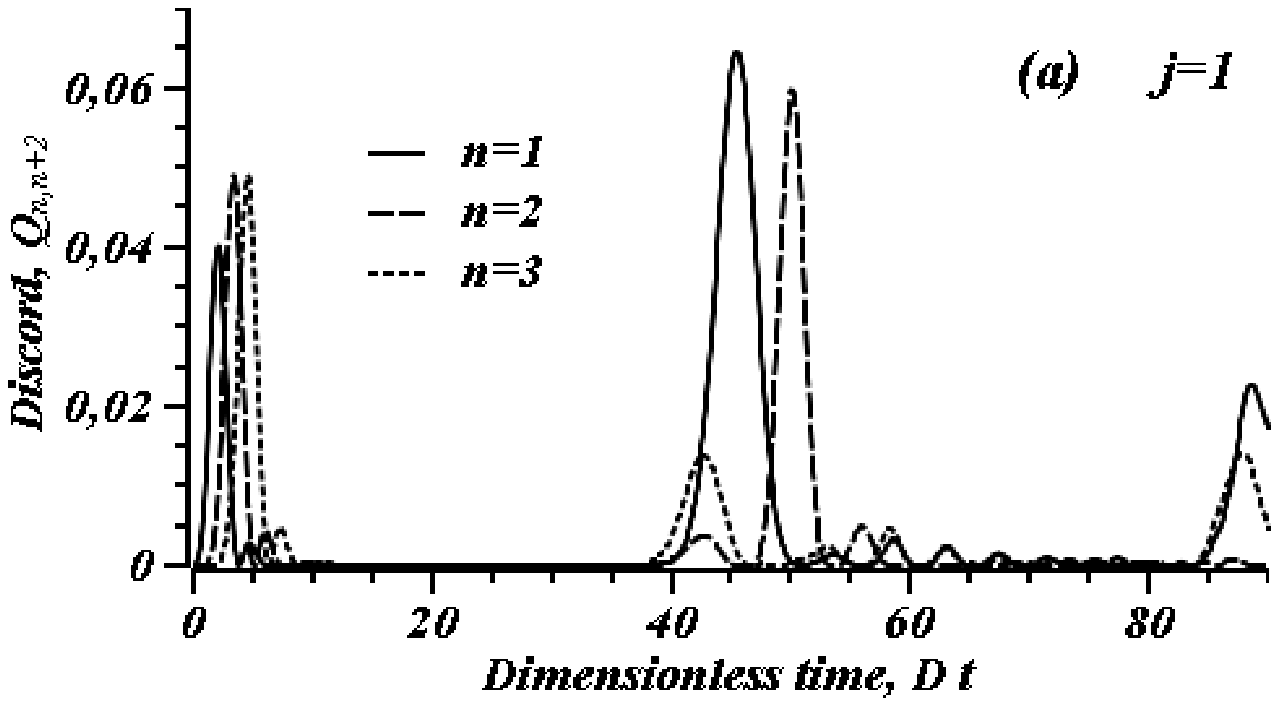
   , scale=0.6,angle=0
}\\
\epsfig{file=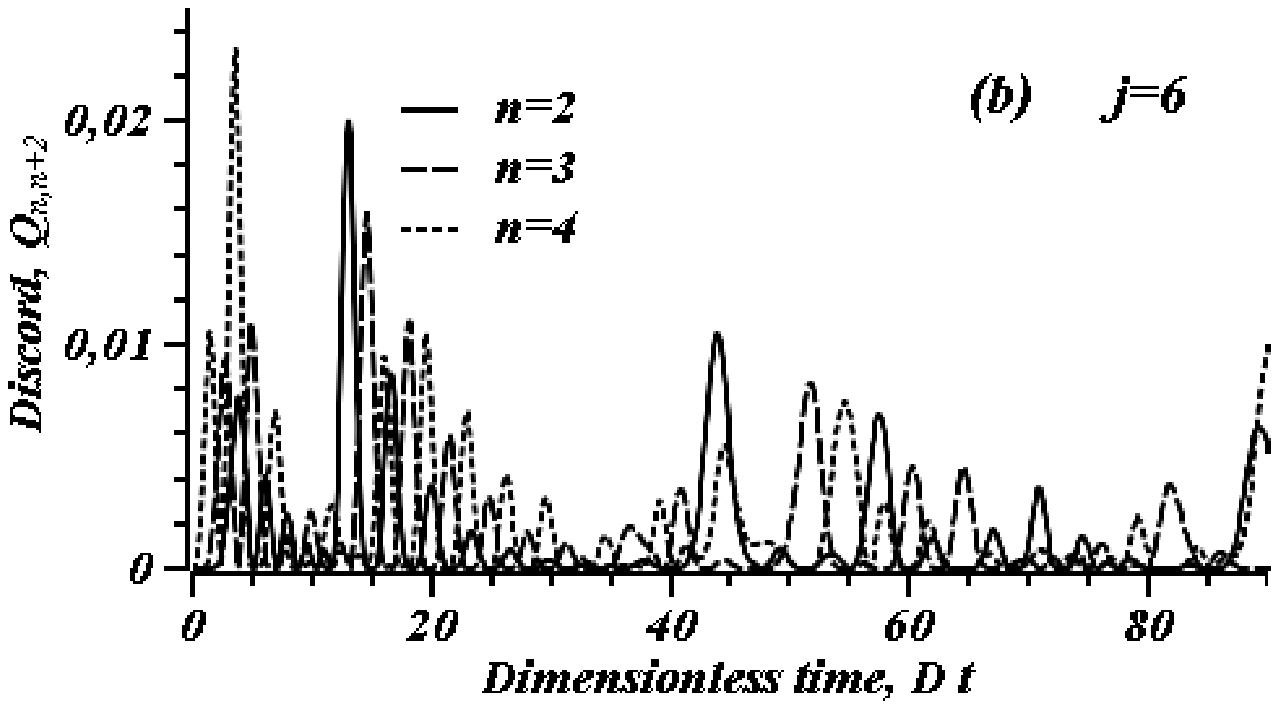
   , scale=0.6,angle=0
}\\
\epsfig{file=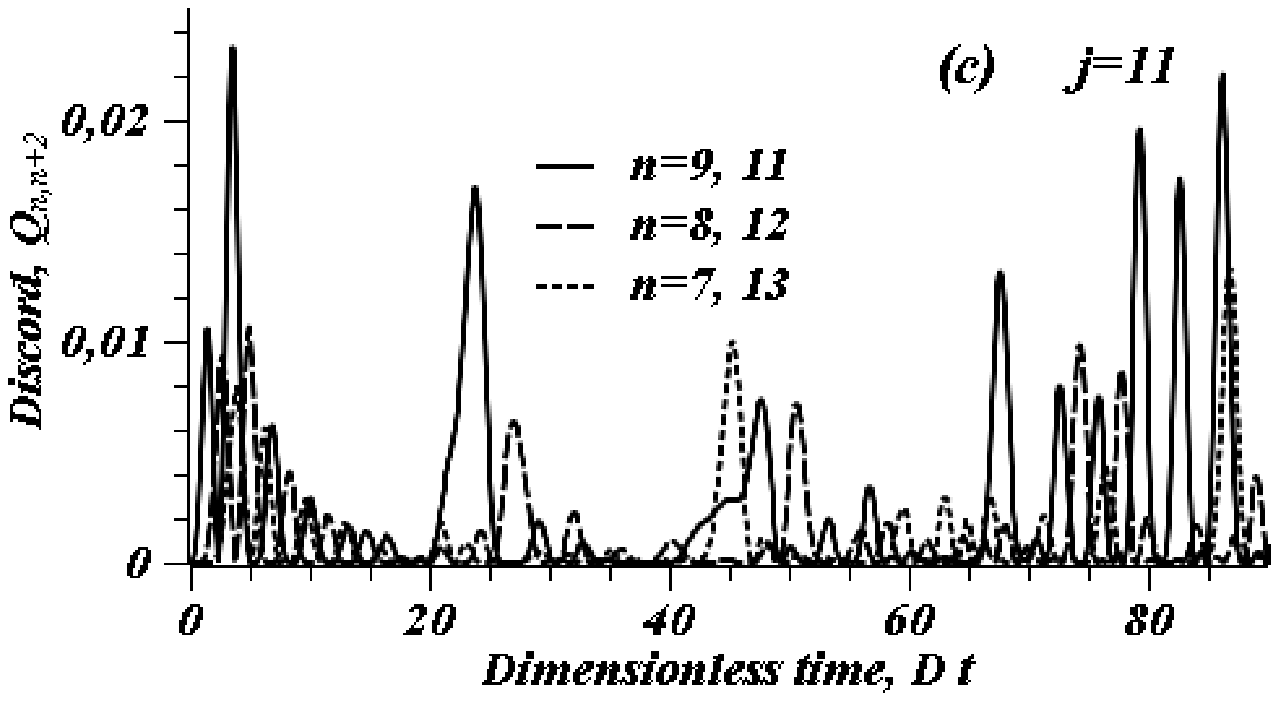
   , scale=0.6,angle=0
}
  \caption{ The time evolution of the discord $Q_{n,n+2}$  for $N=21$ and different initially polarized nodes $j=1,6,11$}
  \label{Fig:c_Disc_pl2} 
\end{figure*}

Now we consider the $c$-representation of the density matrix  
(see Sec.\ref{Section:bas2})
 and  calculate the discord and the concurrence between the $n$th and  $m$th $c$-fermions.  In this case the $\beta$-dependence ($\beta=\frac{\hbar \omega_0}{ k T}$) of the discord is similar to that given in Fig.\ref{Fig:bet} and it will not be discussed here. 
Now the entanglement is non-zero for any $N$ (unlike the entanglement between the $\beta$-fermions). Both the discord and the concurrence evolve in time. This evolution for  $\beta=10$, $N=21$ and different initially polarized nodes $j=1,6,11$ is shown in Fig.\ref{Fig:c_C} (the discord $Q^{c}_{n(n+1)}$ and the concurrence $C^c_{n(n+1)}$ between the nearest neighbors) and in Fig.\ref{Fig:c_Disc_pl2} (the discord $Q^c_{n(n+2)}$).   Emphasize the principal differences between the discord and the concurrence.  
\begin{enumerate}
\item
The concurrence is not zero only between the nearest neighbors (except for the concurrence
$C^c_{13}$, which is not zero in the case of the second initially polarized node, $j=2$; this concurrence is not shown in Fig.\ref{Fig:c_C} ), while the discord between any two  nodes is not identical to 
zero. For instance, the discord $Q^c_{n,n+2}$ is shown in Fig.\ref{Fig:c_Disc_pl2}. 
\item
The concurrence is not zero  only between the neighbors which are closely situated with respect to the initially polarized node.
Thus, if $j=1$, then $C^c_{12}$, $C^c_{23}$ and $C^c_{34}$ are nonzero; if $j=6, 11$, then only two concurrences
(namely, $C^c_{j, j\pm 1}$) are non-zero.
\item
After the discord $Q^c_{nm}$  disappears, it appears again at larger times, see Fig.\ref{Fig:c_C}, unlike the concurrence which does not appear again. We may refer to this phenomenon as the "echo" in the evolution of the discord, see  Fig.\ref{Fig:c_C}a-c (where the "echo" in  evolution of the discord $Q^c_{n(n+1)}$ is shown for different initially polarized nodes ($j=1,6,11$)) and Fig.\ref{Fig:c_Disc_pl2} (where the "echo" in  evolution of the discord $Q^c_{n(n+2)}$ is shown). 
This phenomenon is not observed for the concurrence.
\item
Figs.\ref{Fig:c_C}a-c and \ref{Fig:c_Disc_pl2} demonstrate that the discords $Q_{n(n+l)}$ and $Q_{(n+1)(n+l+1)}$ are both non-zero over some common time interval, which is not observed for the concurrence, see Fig.\ref{Fig:c_C}d-f. This is especially evident for the short time intervals, $t\lesssim 10$. This property of the discord might be used, for instance, to organize the relay of  the quantum discord from the given pair of nodes to the neighboring pair and so on.
\end{enumerate}

\subsection{Discord and concurrence in a system of spin-1/2 particles}
\label{Section:spins}
Now we consider the  basis of eigenvectors of the operators $I_{jz}$ and study the quantum correlations  between the $n$th and $m$th spins (see Sec.\ref{Section:bas3} for the reduced density matrix representation).  
In this case the concurrence is non-zero and both the discord and the concurrence evolve in time. 
Emphasize that  both the concurrence  and the discord are non-zero only between the  nearest neighbors (because the discord and the concurrence in the
system with the diagonal density matrix are zero and $\rho^{spin}_{nm}$ is non-diagonal only if $m=n+1$, see Sec.\ref{Section:bas3}). Moreover, since the reduced  density matrix for the nearest neighbors coincides with that for the $c$-representation  (see Sec.\ref{Section:bas3}), both the discord and the concurrence 
coincide with those obtained in Sec.\ref{Section:c_repr}, i.e  $Q^{{spin}}_{n,n+1} = Q^{c}_{n,n+1}$ and 
$C^{{spin}}_{n,n+1} = C^{c}_{n,n+1}$, see Fig.\ref{Fig:c_C}.  The basic features of the quantum correlations are following.
 \begin{enumerate}
 \item
 Similar to the $c$-representation, 
 the concurrence is non-zero  only between the neighbors which are closely situated  to the initially polarized node.
Thus, if $j=1$, then $C_{12}$, $C_{23}$ and $C_{34}$ are nonzero; if $j=6,11$, then only two concurrences (namely, $C_{j, j\pm 1}$) are not zero.
\item
Similar to Sec.\ref{Section:c_repr}, there is so-called "echo" in the evolution of the discord, while this phenomenon is not observed for the concurrence, see Fig.\ref{Fig:c_C}. 
\item
A principal disadvantage of this matrix representation in comparison with the $c$-representation is the zero discord between remote nodes. 
\end{enumerate}

\section{Conclusions}
\label{Section:conclusions}
For the spin-1/2 open chain of $N\gtrsim 10$ nodes with the XY Hamiltonian we demonstrate  that the behavior of the quantum correlations crucially depends on the basis where these correlations are calculated.
Thus, both the discord  and the concurrence in the $\beta$-representation do not evolve in time  and concurrence is zero in this case.
It is remarkable that the discord is nonzero for the $\beta$-representation reflecting the quantumness of the system. It is shown that the system of $\beta$-fermions with the middle initially excited node possesses  the subsystem of nodes with all equal pairwise discords (the system of odd nodes). If the first node is initially excited, then all nodes are correlated, but the measure of correlations (discord) is different (but non-zero) for each particular pair.

The behavior of the discord and the concurrence in another fermion representation (the $c$-representation) significantly differs from their behavior in the $\beta$-representation. Both the discord and the concurrence  evolve in time, but the concurrence is nonzero only between the nearest neighbors (there is  the single exception for $C^c_{23}$ in the chain with the second initially polarized node), unlike the discord. In addition, the "echo" is observed in the discord evolution, i.e. after the discord $Q_{n,m}$ (with any $m>n$) disappears, it appears again at larger times. In this representation we observe the relay of discord from the first pair of the neighboring nodes to the next pair and so on.  

Finally, the behavior of the concurrence between the nearest neighbors in the basis of eigenvalues of the operators $I_{jz}$ coincides with that for the $c$-representation.
The discord in this basis coincides with the discord in the system of $c$-fermions  only for the nearest neighbors and equals zero (along with the concurrence) between the remote nodes (i.e. $m>n+1$). 

The proposed analysis of the quantum correlations shows that the representation of a quantum system in the basis of eigenvalues of the operators $I_{jz}$   is less applicable in quantum devices in comparison with both fermion  representations. 
Since the  $\beta$- and $c$-representations exhibit more variety of  quantum correlations, it is more profitable to use the fermions  (instead of the real spin-1/2 particles) as nodes in quantum devices.
This example demonstrates that the ''virtual particles'' may reveal some hidden usefull properties  of a given quantum system.

This work is supported by the Program of the Presidium of RAS No.8 ''Development of methods of obtaining chemical compounds and creation of new materials''.


\begin{thebibliography}{99}

\bibitem{Werner}
R.F.Werner, Phys.Rev.A {\bf 40}, 4277 (1989)

\bibitem{HWootters}
S.Hill and W.K.Wootters, Phys. Rev. Lett. {\bf 78}, 5022 (1997)

\bibitem{P}
A.Peres, Phys. Rev. Lett. {\bf 77}, 1413 (1996)


\bibitem{AFOV}
L.Amico, R.Fazio, A.Osterloh and V.Vedral, Rev. Mod. Phys. {\bf 80}, 517 
(2008)

\bibitem{DPF}
S.I.Doronin, A.N.Pyrkov and E.B.Fel'dman, JETP Letters {\bf 85}, 519 (2007)



\bibitem{Z0}
W. H. Zurek, Ann. Phys.(Leipzig), {\bf 9}, 855 (2000)
 

\bibitem{HV}
L.Henderson and V.Vedral J.Phys.A:Math.Gen. {\bf 34}, 6899 (2001)


\bibitem{OZ}
H.Ollivier and W.H.Zurek, Phys.Rev.Lett. {\bf 88}, 017901 (2001) 

\bibitem{Z}
W. H. Zurek, Rev. Mod. Phys. {\bf 75}, 715 (2003)


\bibitem{DSC}
A. Datta, A. Shaji, and C. M. Caves, Phys. Rev. Lett. {\bf 100}, 050502 (2008).




\bibitem{L} 
S.Luo, Phys.Rev.A {\bf 77}, 042303 (2008)

\bibitem{ARA} 
 M.Ali, A.R.P.Rau, G.Alber, Phys.Rev.A {\bf 81}, 042105 (2010)

\bibitem{Xu}
J.-W. Xu, arXiv:1101.3408 [quant-ph]

\bibitem{G}
M.Goldman, {\it Spin temperature and nuclear magnetic resonance in solids}, Oxford, Clarendon Press, 1970


\bibitem{KZ}
 E.I.Kuznetsova and A.I.Zenchuk, Phys.Lett.A {\bf 376},  1029 (2012)


\bibitem{Bose}
S.Bose, Phys.Rev.Lett, {\bf 91} 207901 (2003)





\bibitem{CDEL}
 M.Christandl, N.Datta, A.Ekert and A.J.Landahl, Phys.Rev.Lett. {\bf 92}, 187902 (2004)

\bibitem{ACDE}
 C.Albanese, M.Christandl, N.Datta and A.Ekert, Phys.Rev.Lett. {\bf 93}, 230502 (2004)


\bibitem{KS}
 P.Karbach and J.Stolze, Phys.Rev.A {\bf 72}, 030301(R) (2005)


\bibitem{GKMT}
 G.Gualdi, V.Kostak, I.Marzoli and P.Tombesi, Phys.Rev. A {\bf 78}, 022325 (2008)

 \bibitem{FKZ}
 E.B.Fel'dman, E.I.Kuznetsova and A.I.Zenchuk, Phys.Rev. A {\bf 82}, 022332 (20010)


\bibitem{ZME}
S.Zhang, B.H. Meier and R.R.Ernst, Phys.Rev.Lett. {\bf 69}, 2149 (1992)

\bibitem{FBE}
E.B.Fel'dman, R.Br\"uschweiler and R.R.Ernst, Chem.Phys.Lett. {\bf 294}, 297 (1998)

\bibitem{Z_a}
A.I.Zenchuk, Quant.Inf.Proc. 
DOI 10.1007/s11128-011-0319-x


\bibitem{JW}
P.Jordan, E.Wigner, Z.Phys. {\bf 47}, 631 (1928)


\bibitem{HW}
S.Hill and W.K.Wootters, Phys. Rev. Lett. {\bf 78}, 5022 (1997)

\bibitem{Wootters}
 W. K. Wootters, Phys. Rev. Lett. {\bf 80}, 2245 (1998)


\bibitem{FZ}
E.B.Fel'dman and A.I.Zenchuk, JETP Lett. {\bf 93}, 459 (2011) 













 



\end{thebibliography}
\end{document}